\documentclass[10pt,a4paper,twoside]{article}
\usepackage{epsfig}
\usepackage{baltlat6}
\usepackage{array}
\usepackage{here}
\usepackage{amsmath}
\begin{document}
\ \
\vspace{0.5mm}
\setcounter{page}{1}

\titlehead{Baltic Astronomy, vol.\,XX, 1--7, 2015}

\titleb{ON POSSIBILITY OF APPLYING THE QUASI-ISOTHERMAL ST\"ACKEL'S MODEL TO OUR GALAXY}

\begin{authorl}
\authorb{A. O. Gromov}{1},
\authorb{I. I. Nikiforov}{2},
\authorb{L. P. Ossipkov}{1}
\end{authorl}

\begin{addressl}
\addressb{1}{Department of Space Technologies and Applied Astrodynamics,\\
Saint Petersburg State University, Universitetskij pr.~35,
Staryj Peterhof,\\ Saint Petersburg 198504, Russia; granat08@yandex.ru}
\addressb{2}{Sobolev Astronomical Institute, Saint~Petersburg State University,\\
Universitetskij pr.~28, Staryj Peterhof, Saint Petersburg 198504, Russia;\\ nii@astro.spbu.ru}
\end{addressl}

\submitb{Received: 2014 September 15; accepted: 2015 XXX XX}

\begin{summary}
Earlier the quasi-isothermal St\"ackel's model of mass distribution in stellar systems was
considered. The expression for spatial density was found. In~this work an application of
this model to our Galaxy is studied. The model rotation curve was fitted to data on
kinematics of neutral hydrogen. Estimations of structural and scale parameters of the
model and equidensities for our Galaxy are presented.
\end{summary}


\resthead{Quasi-isothermal model of the Galaxy}
{A. O. Gromov, I. I. Nikiforov, L. P. Ossipkov}

\sectionb{1}{INTRODUCTION}

There are various methods of modelling the gravitational
potential of our Galaxy and other stellar systems. One of them is based
on assumption that the potential is of St\"ackel's form.  After pioneering
works by Kuzmin (1952, 1956) the method was further developed by de Zeeuw et al.\ (1986),
Dejonghe \& de Zeeuw (1988) and others. In this paper we study possibilities
of using the quasi-isothermal potential suggested by Kuzmin et al.\ (1986)
for constructing St\"ackel's model of our Galaxy.

It is known that the third quadratic in velocities integral of motion
\begin{equation}
\label{I3}
I_3=(Rv_z-zv_R)^2+z^2v_\theta^2+z_0^2(v_z^2-2\Phi^*)\,,
\end{equation}
exists for St\"ackel's potentials. Here $R$, $z$ are the cylindrical coordinates; $v_R$,
$v_\theta$, $v_z$ are the projections of spatial velocity; $z_0$ is a scale parameter of dimension of
length; a function $\Phi^*(R,z)$ must satisfy the equations
\begin{equation}
\displaystyle
\displaystyle z_0^2\frac{\partial\Phi^*}{\partial R}=z^2\frac{\partial\Phi}{\partial R}-Rz\frac{\partial\Phi}{\partial z}\,,\qquad
\linebreak
\displaystyle z_0^2\frac{\partial\Phi^*}{\partial z}=(R^2+z_0^2)\frac{\partial\Phi}{\partial z}-Rz\frac{\partial\Phi}{\partial R}\,,
\end{equation}
where $\Phi(R,z)$ is an axisymmetric potential. The assumption on the existence $I_3$ allows to explain the observed triaxiality of
velocity ellipsoid.

In this work we suppose that the potential in the equatorial plane has the following form:
\begin{equation}
\label{Phi}
\Phi(R,0)=\Phi_0\ln\left[1+\frac\beta{w(R)}\right],
\end{equation}
where $\beta\in[0,+\infty)$ is a structural parameter of the model,
\begin{equation}
\label{w2}
w^2(R)=1+\kappa^2 R^2,
\end{equation}
$\Phi_0$ and $\kappa$ are scale parameters.

This potential was proposed by Kuzmin et al.\ (1986) for spherical systems and was called
quasi-isothermal. The potentials by Schuster--Plummer and Jaffe are limiting cases of
this one (when $\beta\rightarrow 0$ and $\beta\rightarrow\infty$ respectively).

St\"ackel's models of mass distribution with the quasi-isothermal potential were
constructed and an analytical (though cumbersome) expression for a spatial density was found
(Gromov 2012, 2013, 2014). St\"ackel's models with the Schuster--Plum\-mer and Jaffe
potentials were also studied (Gromov 2013, 2014).

\sectionb{2}{DATA}

In this study we use data on the rotation of the neutral hydrogen: five independent data
sets from the whole 21-cm line profile and one set from the 21-cm tangent points (see
details in Nikiforov \& Petrovskaya 1994). The whole profile data contain
Camm's function values $\Omega\equiv R_0(\omega
-\omega_{\rm LSR})$ in relation to $x\equiv R/R_0$, where $R$ is the distance to the galactic axis,
$\omega$ and $\omega_{\rm LSR}$ are the angular velocities of H\,I at the distance $R$ and
of the Local Standard of Rest respectively, $R_0$ is the distance of the Sun to the
Galactic center. The tangent points' data also define in fact the dependence of $\Omega$
on $x$. The total number of H\,I data points is 239.

It should be pointed out that analyses of the whole 21-cm profiles give at any radius the
velocity of rotation averaged over essentially all galactocentric circle of this radius.
Therefore the resultant H\,I rotation curve is well smoothed over essentially all
galactocentric longitudes.

A value of the linear velocity of rotation, $v_i$, for an H\,I data point $(x_i,\Omega_i)$
can be calculated by the formula
\begin{equation}
\label{vi}
v_i=\left[\Omega_i+v_{\rm LSR}\right]x_i,
\end{equation}
where $v_{\rm LSR}$ is the linear velocity of the LSR. Here we adopted a value for $v_{\rm
LSR}=220$~km/s as a combination of constants $v_{\rm LSR}=R_0\,\omega_{\rm LSR}$, where
$R_0=8.0$~kpc (e.g., Reid 1993; Nikiforov 2004; Nikiforov \& Smirnova 2013), $\omega_{\rm
LSR}=27.5$~km\,s$^{-1}$\,kpc$^{-1}$ is an intermediate value between recent determinations
(e.g., Feast \& Whitelock 1997; Zabolotskikh et al.\ 2002). Since as a rule the number of
points in a H\,I data set can be arbitrary set by authors of study, the weights of data
points, $p_i$, were adopted to be proportional to the length of interval $[x_{\text{min}},
x_{\text{max}}]$ covered by this set (see Table~1 in Nikiforov \& Petrovskaya 1994).

Notice that the form of the rotation curve based on the H\,I data is independent of
calibrations of distance scales and is specified by only the parameter $v_{\rm LSR}$,
according to Eq.~(\ref{vi}). Choosing a value for $R_0$ affects the value of $v_{\rm
LSR}$, with our parametrization of the H\,I rotation curve, and scales the
coefficient~$\kappa$ in Eq.~(\ref{w2}), in any case. Other parameters of the model
potential~(\ref{Phi}) do not directly depend on an adopted value of $R_0$.

\begin{figure}[!tH]
\vbox{
\centerline{\psfig{figure=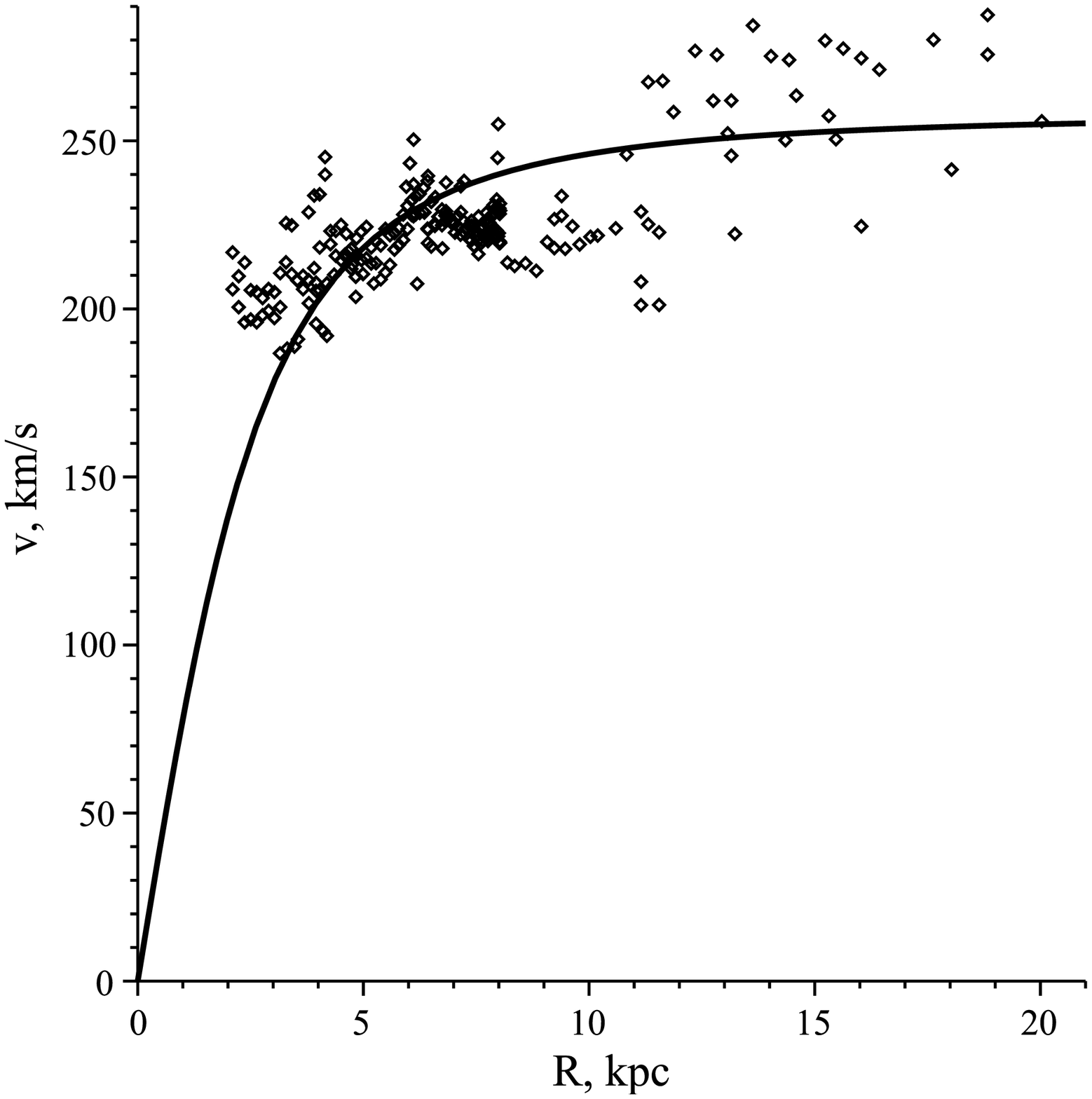,width=90mm,angle=0,clip=}}
\vspace{1mm}
\captionb{1}
{Comparison of the model circular velocity curve with data on the rotation of the neutral hydrogen.}}
\end{figure}

\sectionb{3}{ESTIMATION OF $\Phi_0$, $\beta$, $\kappa$}

To construct the St\"ackel's model potential~(\ref{Phi}), (\ref{w2}) for our Galaxy
it is necessary to estimate parameters $\Phi_0$, $\beta$, $\kappa$. We shall fit the
model circular velocities, $v_{\text{c}}$, to observational data on the rotation of
the Galaxy. An expression for the circular velocity
\begin{equation}
v_{\text{c}}^2(R)=-R\frac{\partial\Phi}{\partial R}(R,0)
\end{equation}
in the case of the quasi-isothermal potential has the following form:
\begin{equation}
\label{vc2}
v_{\text{c}}^2(R)=\Phi_0\frac{\beta\kappa^2R^2}{\left(1+\kappa^2R^2\right)^{\frac32}\left(1+\displaystyle\frac{\beta}{\sqrt{1+\kappa^2R^2}}\right)}\,,
\end{equation}
where
$q=\displaystyle\frac{\beta}{\beta+1}$. We shall use $q$ as a structural parameter
instead of $\beta$ for $q\in [0,1]$.

The model parameters were estimated by the ordinary least-squares fitting.
We minimized the statistics
\begin{equation}
L^2=\sum_{i=1}^{239}p_i\left[v_{\text{c}}(R_i)-v_i\right]^2,
\end{equation}
where $v_{\text{c}}(R_i)$ is the model value of circular velocity at $R_i$ calculated
by Eq.~(\ref{vc2}), $v_i$ is the ``observed'' value of circular velocity
calculated by Eq.~(\ref{vi}), $p_i$ is the weight.

We found, that minimum $L^2$ is achieved, when $\Phi_0=258.1\pm1.5$ km$^2$\,s$^{-2}$,
$\kappa=(0.32\pm0.01)\left({R_0/8\text{ kpc}}\right)^{-1}$~kpc$^{-1}$, $\displaystyle
q=1^{+0}_{-0.008}$\,. With these values, the quasi-isothermal model gives the best
approximation for observational data.

A comparison of the model curve of circular velocity with observational data is shown on
Fig. 1, where the solid curve is the model velocity curve $v_{\text{c}}(R)$ and points are
the H\,I data. The mean error of weight unit for this solution is
$\sigma=2.98$~km\,s$^{-1}$. Although a polynomial model of rotation curve can reveal a
finer structure of the H\,I rotation law with $\sigma=2.10$~km\,s$^{-1}$ for the same data (Nikiforov
2000), Fig.~1 demonstrates that the quasi-isothermal potential in the equatorial plane
represents the main trend in the rotation low and can be used for an approximation of the
rotation curve for our Galaxy.

It is worth noting that the form of rotation curve for the quasi-isothermal potential
(Fig.~1) is similar to the form of the ``universal'' rotation curve for spiral galaxies
(Persic et al.\ 1996). Recent results of kinematic modelling the Galactic subsystem of
high-mass star forming regions (Reid et al.\ 2014) demonstrate that the universal rotation
curve of Persic et al.~(1996) is well suited to the representation of the rotation curve
of the Galaxy.


\sectionb{4}{ESTIMATION OF $z_0$}

The expression for density of St\"ackel's models includes the parameter $z_0$. But it
is impossible to find the latter from the rotation curve. In principle, $z_0$ can be
estimated from the run of total spatial density, but presently we cannot do it for
any one-component St\"ackel's model. Indeed, we do not know the density run for the
dark matter from observations. Hence, to estimate $z_0$ we are obliged to suppose
that in the Solar neighborhood the potential of our model is close to potentials of
other models constructed with using not only the rotation curve but other
observational data as well. The model by Gardner et al.\ (2011)  is the last of such
models. Its potential is
\begin{equation} \label{9}
\Phi=\Phi_H+\Phi_C+\Phi_D+\Phi_g\,,
\end{equation}
where
$$
\Phi_H=\frac12 V_h^2 \ln(R^2+z^2+R_1^2),
$$
$$
\Phi_C=-\frac{GM_{C_1}}{\sqrt{R^2+z^2+R_{C_1}^2}}-\frac{GM_{C_2}}{\sqrt{R^2+z^2+R_{C_2}^2}}\,,
$$
$$
\Phi_D=\sum_{i=1}^3\frac{-GM_{d_i}}{\sqrt{R^2+\left(a_{d_i}+\sqrt{z^2+b^2}\right)^2}}\,,
$$
$$
\Phi_g=\sum_{n=1}^3\frac{-GM_{g_n}}{\sqrt{R^2+\left(a_{d_n}+\sqrt{z^2+b^2_g}\right)^2}}\,,
$$ where $G$ is the gravitational constant, the values of parameters
$V_h^2$, $R_1$, $M_{C_1}$, $M_{C_2}$, $R_{C_1}$, $R_{C_2}$, $M_{d_i}$, $a_{d_i}$, $b$, $M_{g_n}$, $b_g$ are given by Gardner et al.\
(2011).

\begin{figure}[!tH]
\vbox{
\centerline{\psfig{figure=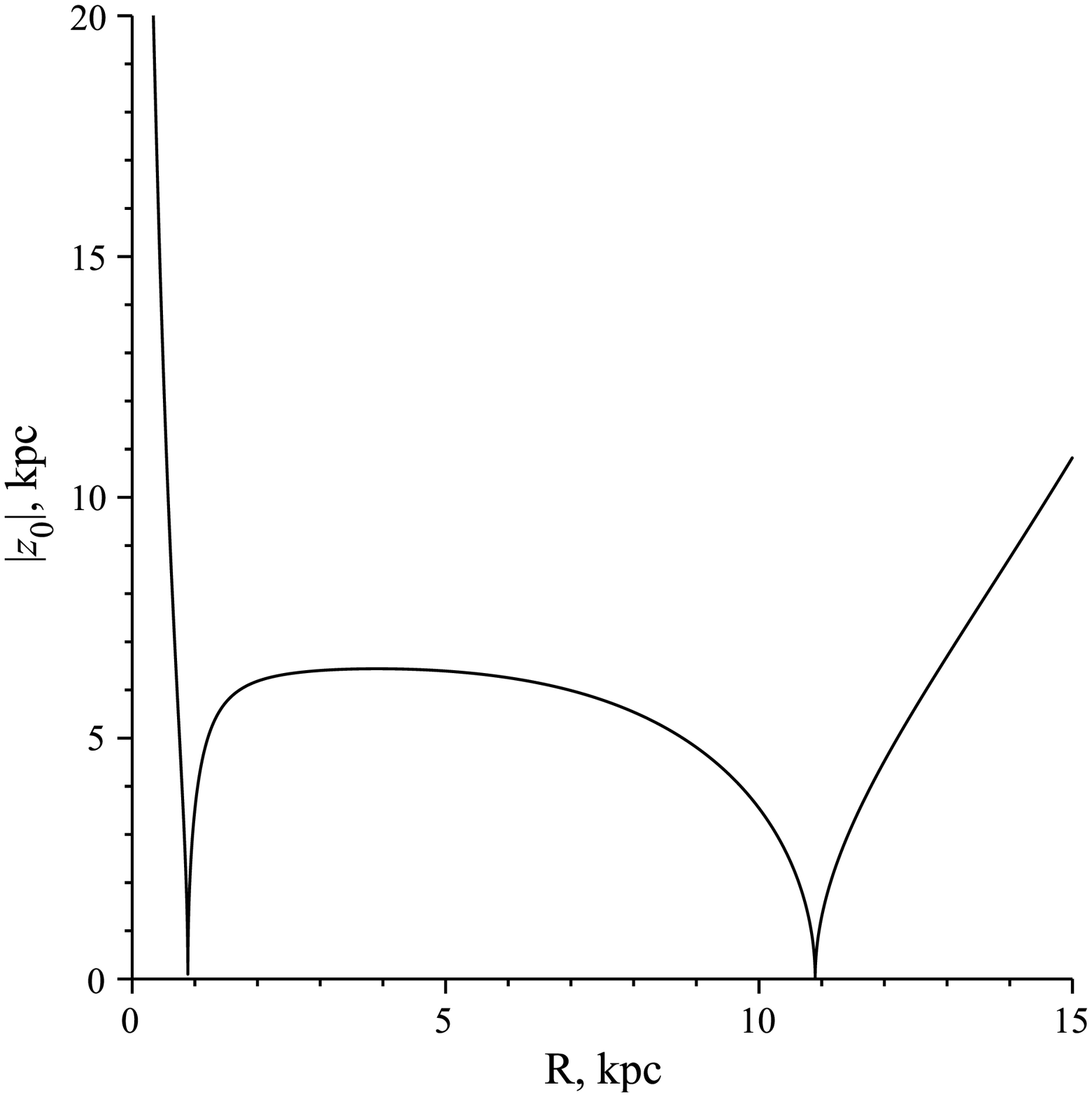,width=90mm,angle=0,clip=}}
\vspace{1mm}
\captionb{2}
{The function $|z_0(R)|$ for the model of Gardner et al.\ (2011).}
}
\end{figure}

For any potential $\Phi (R, z)$ it is possible to define a function
\begin{equation}
\displaystyle
z_0^2(R)=\left.\displaystyle\left[\frac{3\,\displaystyle\frac{\partial \Phi(R,z)}{\partial R}+R\left(\displaystyle\frac{\partial^2 \Phi(R,z)}{\partial R^2}-4\,\frac{\partial^2 \Phi(R,z)}{\partial z^2}\right)}{\displaystyle\frac{\partial^3 \Phi(R,z)}{\partial  z ^2 \partial R}}\right]\right|_{z=0}-R^2
\end{equation}
(Kuzmin 1952, Ossipkov 1975). It follows from the condition of existence of the third
quadratic  integral (\ref{I3}) that  $z_0^2(R)\equiv\rm {const}$ for St\"ackel's
models. Fig.~2 shows $|z_0(R)|$  for the model of Gardner et al.\ (2011). We see that
this function is almost constant for $R$ from 1.5~kpc to 9~kpc. The latter means that
for such potential the expression (\ref{I3}) can be used as an approximate integral  of
motion for stars moving inside this zone (with not very large $|v_z|$). Functions
$z_0(R)$ were constructed earlier by Einasto \& R\"ummel (1970) and Ossipkov (1975)
for Einasto's models of our Galaxy and M~31. These authors discussed the negativity of
$z_0^2(R)$ for large $R$ that takes place also for the potential (\ref{9}).

We conclude that it is possible to set $z_0=5.4$~kpc (this is a value of $|z_0|$ at
$R_0=8$~kpc for the model of Gardner et al.\ 2011). Earlier Ossipkov (1975) found
$z_0=7$~kpc, and according to Kuzmin (1956) $z_0=3.6$~kpc. Our value of $z_0$ is
close to $z_0=4.8$~kpc obtained by Malasidze (1973).

\sectionb{5}{MODEL OF THE GALAXY WITH THE QUASI-ISOTHERMAL\\ POTENTIAL}

For constructing the quasi-isothermal model of Galaxy the obtained values of parameters
were substituted in an expression for spatial density (Gromov 2012, 2013, 2014). As a
result, the equidensities  for central layers of Galaxy were obtained. They are shown on
Fig. 3. As earlier (Gromov 2012, 2013, 2014), $\rho_0$ is a density in the center of the
model, when $\kappa=1 $~kpc$^{-1}$, $\beta=1$, $\Phi_0=1$ km$^2$\,s$^{-2}$.

\begin{figure}[!tH]
\vbox{
\centerline{\psfig{figure=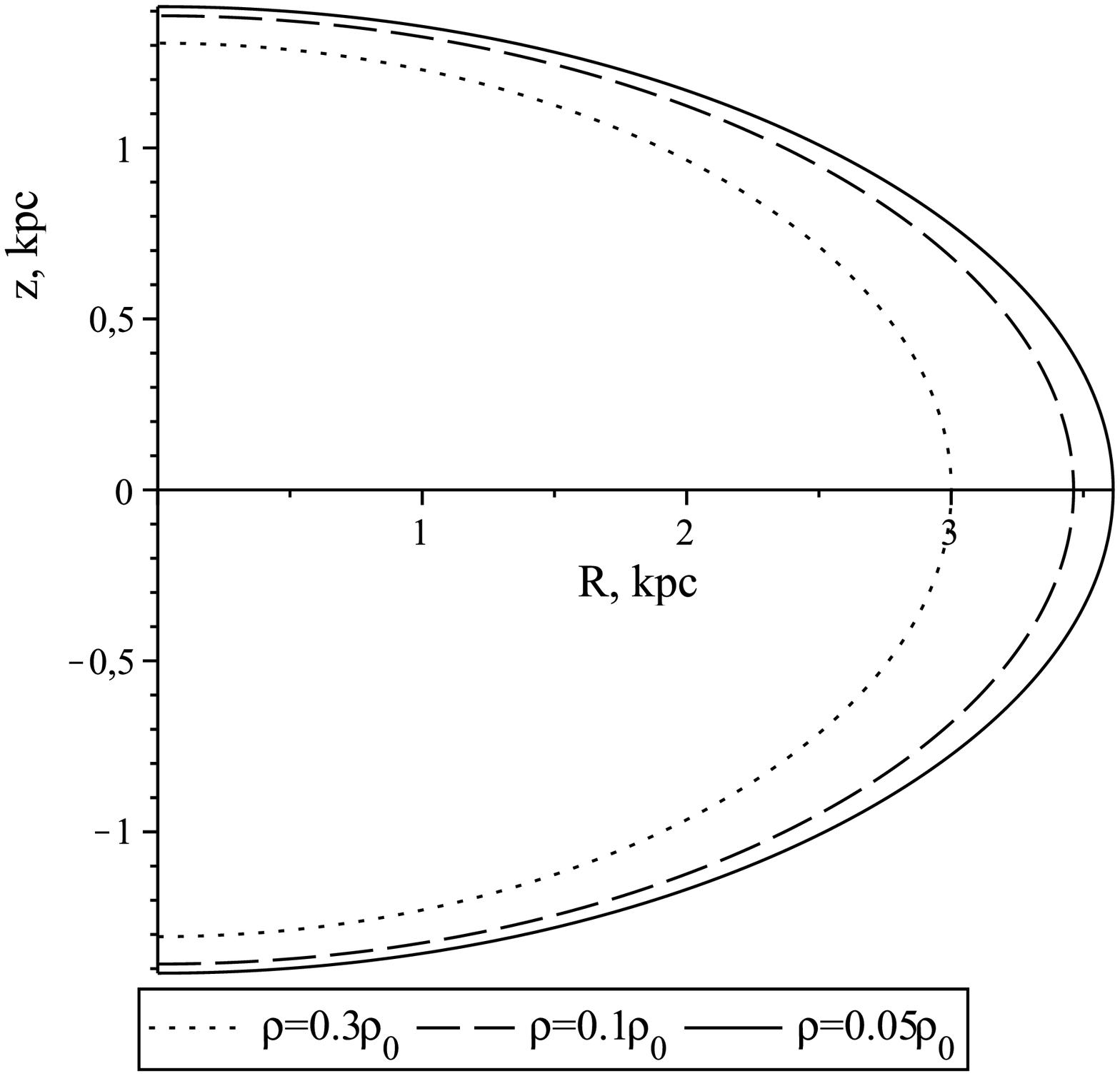,width=90mm,angle=0,clip=}}
\vspace{1mm}
\captionb{3}
{Equidensities for obtained values of parameters.}
}
\end{figure}

These equidensities are similar to equidensities for Jaffe's model,
which is the limiting case ($\beta\rightarrow\infty$) of quasi-isothermal model
\begin{equation}
\Phi(R,0)=\Phi_0\ln\left(1+\lambda^{-1}R^{-1}\right),
\end{equation}
where $\lambda$ is a scale parameter.

Fig.~3 shows a drastic drop in density at equidensities which dimensions are close to
ones of the Galactic triaxial bulge/bar---($3.5:1.4:1.0$)~kpc (see Gardner \& Flynn
2010). It is possible that this resemblance is accidental. However, the direct
inclusion of bulge component in the dynamic model with the use of the same H\,I data
(Nikiforov 2001) gives a similar result: a severely limited from above estimate of
the cut-off radius of bulge $a_{\text{b}}=2.1^{+0.6}_{-2.1}$~kpc.

It is evident that the total mass of the quasi-isothermal model is
\begin{equation}
M=\Phi_0\frac{\beta}{\kappa G}\,.
\end{equation}
The minimum value of $M$ for the obtained values of parameters is
$6\times10^{12}\,M_{\odot}$\,. The maximum value of the mass for the same parameter values is
unlimited, because $\beta$ can take infinity large values.

\sectionb{6}{CONCLUSIONS}

We estimated parameters of the quasi-isothermal model of mass distribution for our
Galaxy using data on the rotation of the neutral hydrogen and obtained equidensities
for this set of parameters.

The comparison of the model with observational data shows the relevance of the model.
So, the quasi-isothermal model can be used for approximation of the gravitational
potential of our Galaxy. It is possible to apply the quasi-isothermal model to
external galaxies with almost flat rotation curve.

In the future a multi-component St\"ackel's model  will be constructed for the Galaxy
to gain a better approximation to the data.

\References

\refb Dejonghe H., de Zeeuw T. 1988, AJ, 333, 90

\refb de Zeeuw T., Peletier R., Franx M. 1986, MNRAS, 221, 1001

\refb Einasto J., R\"ummel U. 1970, Astrophysics, 6, 120

\refb Feast M. W., Whitelock P. 1997, MNRAS, 291, 683

\refb Gardner E., Flynn C. 2010, MNRAS, 405, 545

\refb Gardner E., Nurmi P., Flynn C., Mikkola S. 2011, MNRAS, 411, 947

\refb Gromov A. O. 2012, Astron. Tsirkulyar, 1579, 1

\refb Gromov A. O. 2013, Izv.\ glavn.\ astron.\ obs. (Pulkovo), 221, 129

\refb Gromov A. O. 2014, Vest.\ Saint Petersburg Univ., ser. 1, 2, 322

\refb Kuzmin G. G. 1952, Publ.\ Tartu Obs., 32, 332

\refb Kuzmin G. G. 1956, AZh, 33, 27

\refb Kuzmin G. G., Veltmann \"U.-I. K.,  Tenjes P. L. 1986, Publ.\ Tartu Obs., 51, 232


\refb Malasidze G. A. 1973, in  Dynamics of Galaxies and Star Clusters,  ed. T. B.
Omarov, Alma-Ata, Nauka, p.~93

\refb Nikiforov I. I. 2000, in Small Galaxy Groups (IAU Colloquium 174),
    eds. M.~J.~Valtonen \& C.~Flynn, San Francisco, CA, p.~403

\refb Nikiforov I. I. 2001, in Stellar Dynamics: from Classic to Modern,
    eds. L.~P.~Ossipkov~L.P. \& I.~I.~Nikiforov, St.~Petersburg University Press, p.~28

\refb Nikiforov I. I. 2004, in Order and Chaos in Stellar and Planetary Systems,
    eds. G.~G.~Byrd, K. V. Kholshevnikov, A. A. Myll\"ari, I. I. Nikiforov,
    \& V. V. Orlov,  ASP Conf. Ser., 316, 199

\refb Nikiforov I. I., Petrovskaya I. V. 1994, ARep, 71, 725

\refb Nikiforov I. I., Smirnova O. V. 2013, AN, 334, 749

\refb Ossipkov L. P. 1975, Vest. Leningrad Univ., 13, 142

\refb Persic M., Salucci P., Stel  F. 1996, MNRAS, 281, 27

\refb Reid M. J. 1993, ARA\&A,~31, 345

\refb Reid M. J., Menten K. M., Brunthaler A., Zheng X. W., Dame T. M., Xu Y., Wu
Y., Zhang B., Sanna A., Sato M., Hachisuka K., Choi Y. K., Immer K.,
Moscadelli L., Rygl K. L. J., Bartkiewicz, A.
2014, ApJ, 783, 130

\refb Zabolotskikh M. V., Rastorguev A. S., Dambis A. K. 2002, ALett, 28, 454

\end{document}